\documentclass[twocolumn,twoside,slac_two]{revtex4}
\usepackage{graphicx}
\usepackage{fancyhdr}
\newcommand{ \be }{\begin{eqnarray}}       
\newcommand{ \ee }{\end{eqnarray}}       
\newcommand{ \la }{\langle}       
\newcommand{ \ra }{\rangle}

\newcommand{ \mean }[1]{\left\langle #1 \right\rangle}   

\def\snn{$\sqrt{s_{NN}}$}

\def\P{$\cal P$}

\newcommand{ \psirp }{\Psi_{RP}}
\newcommand{ \phia }{\phi_{\alpha}}
\newcommand{ \phib }{\phi_{\beta}}
   
\usepackage{color}   
\definecolor{orange}{cmyk}{0.,0.353,1.,0.}    
\definecolor{dgreen}{cmyk}{1.,0.,1.,0.4}	

\usepackage{fixmath} 
\usepackage{dcolumn} 
\usepackage{bm} 

\begin{document}

\title{
Azimuthal charged particle correlations
as a probe for local strong parity violation
in heavy-ion collisions}

\author{Ilya Selyuzhenkov for the STAR Collaboration}

\affiliation{Indiana University Cyclotron Facility, 2401 Milo B. Sampson Lane, Bloomington, IN 47408, USA}

\begin{abstract}
One of the most interesting and important phenomena predicted
to occur in heavy-ion collisions is the local strong parity violation.
In non-central collisions, it is expected to result in charge separation
of produced particles along the system's orbital momentum.
I report on results of the charge separation measurement
in Au+Au and Cu+Cu collisions at $\sqrt{s_{NN}}= 200$ and 62 GeV
with the STAR detector at RHIC based on
three-particle mixed harmonic azimuthal correlations.
Systematic study of parity conserving (background) effects
with existing heavy-ion event generators,
and their possible contributions to the observed
correlations are also present.
\end{abstract}

\maketitle

\thispagestyle{fancy}

\section{Introduction}

\begin{figure}[ht]
  \includegraphics[width=.42\textwidth]{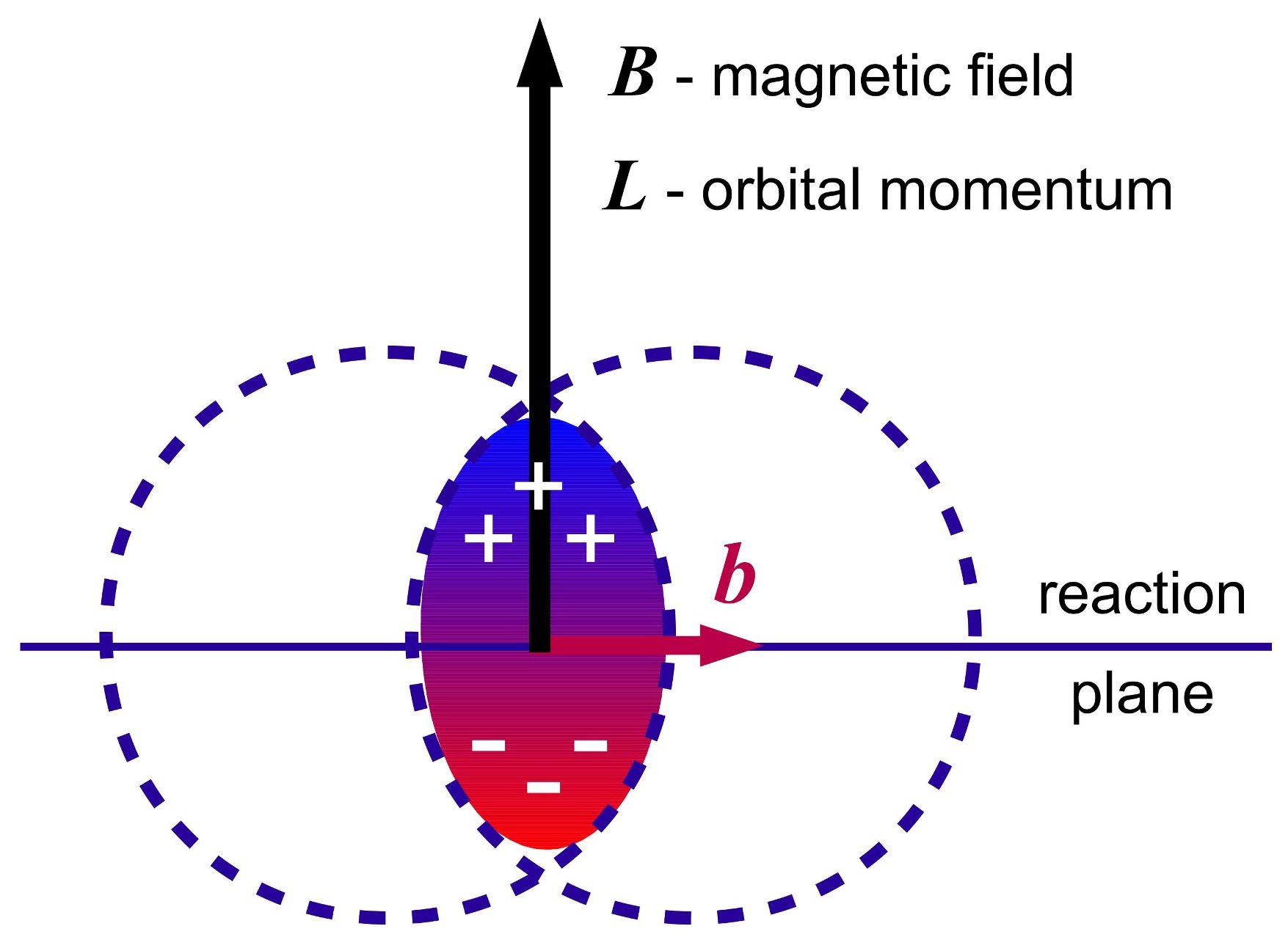}
  \caption{Schematic view of non-central heavy-ion collision.
Colliding nuclei (depicted as dashed circles) are moving out-of-list.
Magnetic field ($B$) and system orbital momentum ($L$) are perpendicular
to the reaction plane (plane spanned by the impact parameter, $b$, and colliding nuclei direction).
}
  \label{fig:orbMom_MagField}
\end{figure}

A heavy-ion collision provides a unique environment
to study particle interactions
at a very high temperature and extreme density.
Among the most important features of the system
created in non-central heavy-ion collisions
are the strong magnetic field
$B \sim 10^{15}$ T ($eB \sim 10^{4}$ MeV${}^2$)~\cite{Kharzeev:2004ey,Kharzeev:2007jp,Fukushima:2008xe},
and large orbital angular momentum $L \sim 10^{5}$~\cite{Liang:2004ph, Gao:2007bc}.
Together with non-uniform particle density and pressure gradient
in the overlap area formed by the colliding nuclei
(depicted by the color region in Fig.~\ref{fig:orbMom_MagField}),
such initial conditions result in a variety
of interesting physics phenomena which are currently under study
at the Relativistic Heavy Ion Collider (RHIC).
Among the most extensively studied is the anisotropic transverse flow,
which reflects the evolution
of initial space anisotropy of particle production
in the overlap area into momentum space.
This collective effect has been measured at RHIC
and SPS (Super Proton Synchrotron)
(see \cite{Voloshin:2008dg} and references therein),
and the strong elliptic and directed flow observed at RHIC are
key measures which indicate
the quark-gluon plasma formation in heavy-ion collisions.

Another interesting phenomena predicted to occur in non-central heavy-ion collisions
is global polarization and spin alignment \cite{Voloshin:2004ha,Liang:2004ph}.
It manifests itself in preferential orientation of
the spin of produced particles wrt. the system orbital momentum
(perpendicular to the reaction plane as illustrated in Fig.~\ref{fig:orbMom_MagField}).
This effect has been recently measured by the STAR
Collaboration
for strange hyperons ($\Lambda$, $\bar \Lambda$)
and vector mesons ($K^{*0}$, $\phi$),
and is experimentally found to be consistent with zero
within experimental uncertainties
\cite{Abelev:2008ica,Abelev:2007zk}.

In this presentation I report on results of the STAR studies
of the local strong parity (\P) violation.
Experimental evidence for this effect would be
an observation of charge separation 
along the direction of the magnetic field
(preferential emission of the same charged particles
in the same direction).
This effect is illustrated
by "+" (positive) and "--" (negative) charge separation in Fig.~\ref{fig:orbMom_MagField}.

\begin{figure}[ht]
  \includegraphics[width=.42\textwidth]{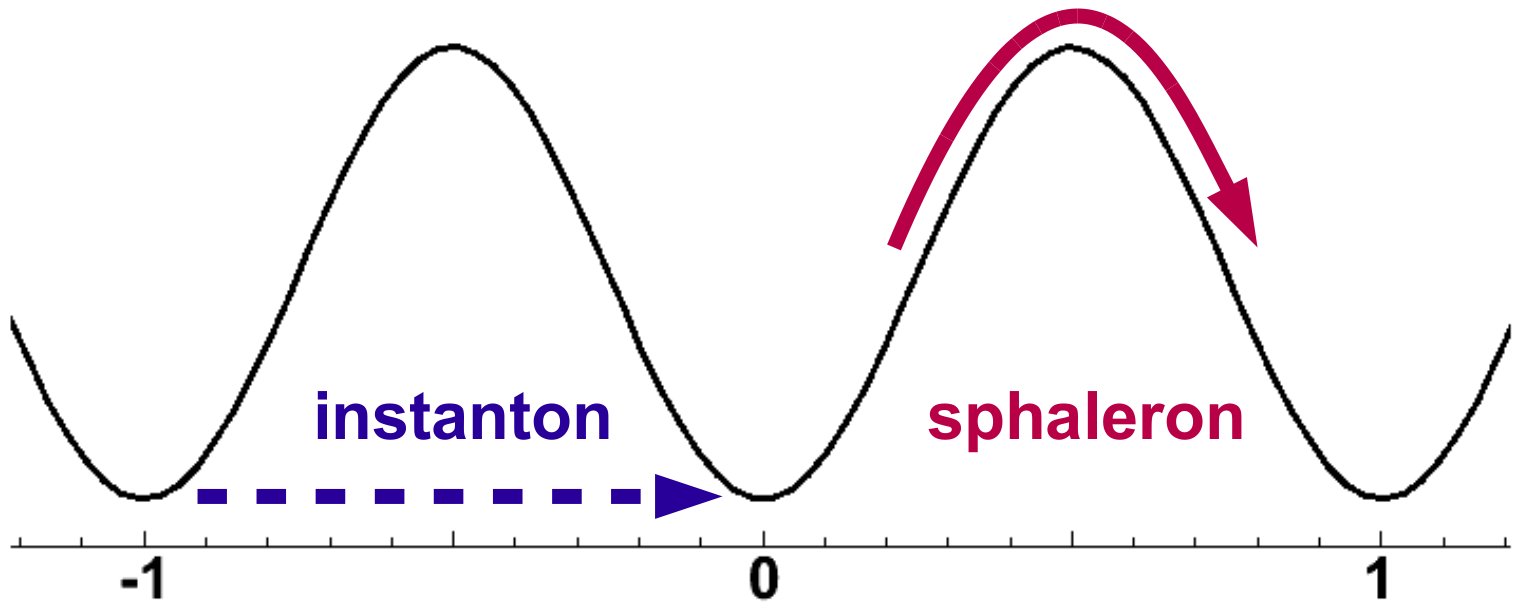}
 \centerline{$N_{CS}$, generalized coordinate}
  \caption{Schematic view of QCD vacuum and its topological transitions
from one local minima to another
(tunneling via instanton, or go-over-barrier via sphaleron).
}
  \label{fig:TopolChargeTransition}
\end{figure}

The underlying physics of charge separation originates in the
fundamental features of the QCD vacuum.
The modern understanding of the QCD theory is that
its vacuum (gluonic field energy) is
periodic vs. so-called Chern-Simons number $N_{CS}$ (related to the
topological charge, $Q$, as $Q= N_{CS}(+\infty) -N_{CS}(-\infty)$) \cite{Diakonov:2009jq},
and there exist localized in space and time solutions
(topological configurations) which
corresponds to the transformation from one local minima to another.
Transition between different vacua
can be either via tunneling (instanton) or go-over-barrier (sphaleron)
(see \cite{Fukushima:2008xe} and references therein).
Figure~\ref{fig:TopolChargeTransition} schematically illustrates these transitions.
Quark interaction with either of these topological configurations
changes chirality, which is a \P~and time reversal odd transition.

At the same time it is experimentally known that
\P~and $CP$ invariance are (globally)
preserved in strong interactions.
The evidence comes form neutron
electric dipole moment experiments,
which put a strong constrains on the 
possible magnitude of the parity violating parameter \cite{Pospelov:1999ha,Baker:2006ts}:
$\theta < 10^{-10}$
(if  $\theta \ne 0$, then QCD vacuum breaks \P~and $CP$ symmetry).

It was recently argued by Kharzeev {\it et. al.}~\cite{Kharzeev:2004ey,Kharzeev:2007jp,Fukushima:2008xe}
that in non-central heavy ion collisions (local) formation of
metastable \P-odd domains is not forbidden.
Such local strong parity violation can be viewed as
a combined result of the following physics effects ~\cite{Fukushima:2008xe}:
\begin{itemize}
\item
External magnetic field aligns quark spins
along or opposite to its direction.
Right-handed quark momentum becomes
opposite to the left-handed one.
\item
Vacuum topological transitions produce a domain with local
excess of left or right handed quarks $N_{\rm left} \ne N_{\rm right}$
(what corresponds to non-zero topological charge).
\item
Magnetic field induces electric field which is parallel to it: $E \sim \theta \cdot B$.
In this electric field positive and negative charges
start to move opposite to each other,
which in a finite volume results in charge separation.
\end{itemize}
Depending on the sign of the domain's topological charge, 
positively charged particles will be preferentially emitted 
either along, or in the direction opposite to, the system orbital 
angular momentum, with negative particles
flowing oppositely to the positive particles.

\section{Experimental observable}
\label{sec:method}
                                  
Charge separation can be described 
by adding \P-odd sine terms to the Fourier decomposition of 
the particle azimuthal distribution 
with respect to the reaction plane angle, $\Psi_{RP}$~\cite{Voloshin:2004vk}: 
\be
 \frac{dN_\pm}{d\phi_\alpha} &\propto& 1 + 2 \sum_{n=1}^{\infty} v_{n,\pm} \cos n \Delta \phi_\alpha
+2 a_{\pm} \sin\Delta \phi_\alpha,
\label{eq:expansion}
\ee
where $\Delta \phi_\alpha =\phi_\alpha-\psirp$ is the particle azimuth 
relative to the reaction plane,
$v_n$ are coefficients accounting for the anisotropic flow
($v_1$ is called  directed, and $v_2$ is elliptic flow).
The $a$ parameters, $a_- = -a_+$, describe the \P-violating effects
(in the present study we consider only the first harmonic).

\begin{figure}[ht]
  \includegraphics[width=.44\textwidth]{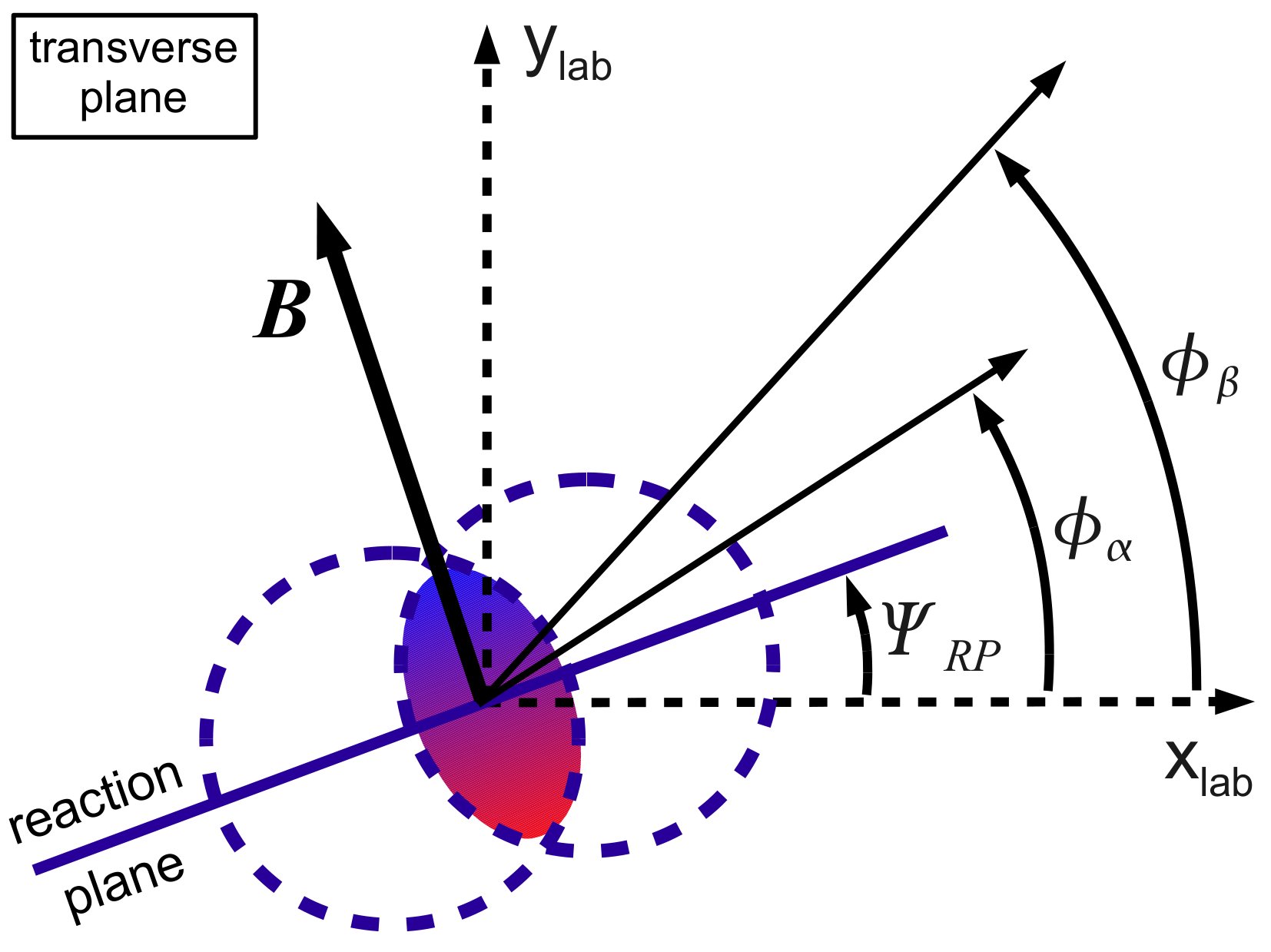}
  \caption{Relation between azimuthal angles of the particle $\alpha$ and $\beta$
in the laboratory frame and the reaction plane.
}
  \label{fig:charge_sep}
\end{figure}

For mid-central collisions at RHIC energies,
predicted value for the $a$ parameter is of the order of 1\% \cite{Kharzeev:2004ey},
which is too small to be observed in a single event.
The direct, \P-odd, observable $\mean{a_\pm}$ (average asymmetry over many events) must also yield zero,
and the only way to experimentally probe
charge asymmetry is to use
multi- (two or more) particle correlations,
what limits the measurement to \P-even observables only.
This brings a problem of separating (or suppressing)
\P-conserving background physics effects  in the measurement.
Physics backgrounds can be sorted into two different categories:
background which produce correlations independent of the reaction plane orientation
and those which are reaction plane dependent.
To suppress reaction plane independent backgrounds, Voloshin proposed to
use the following two particle correlator
wrt. the reaction plane ~\cite{Voloshin:2004vk}:
\be
\hspace*{-2cm}
& \mean{ \cos(\phia +\phib -2\psirp) } = 
\label{eq:obs1}
\\
&=
\mean{\cos\Delta \phia\, \cos\Delta \phib} 
-\mean{\sin\Delta \phia\,\sin\Delta \phib}
\label{eq:cossin}
\\ 
& =
[\mean{v_{1,\alpha}v_{1,\beta}} + B_{in}] - [\mean{a_\alpha a_\beta}
+ B_{out}].
\label{eq:v-a}
\ee
Here $B_{in}$ and $B_{out}$ correspond to the in- and out-of plane
background contributions to the correlator,
and $\alpha,\beta$ denote particle charge.
Figure \ref{fig:charge_sep} illustrates the relation
between various azimuthal angles in the laboratory frame
and the reaction plane orientation.

The correlator (\ref{eq:cossin}) represents the
difference between correlations projected onto an axis in
the reaction plane and the correlations projected onto an axis
that is out-of-plane or perpendicular to the reaction plane. 
The key advantage of using a difference equation is that it
removes all the background correlations among
particles $\alpha$ and $\beta$ that are not related to the reaction plane 
orientation~\cite{Borghini:2002vp,Adams:2003zg}.

\section{Expectations for charge correlations}

The first estimates~\cite{Kharzeev:2004ey}
of the charge asymmetry predicted a signal of the order
 of $|a|\sim Q/N_{\pi^+}$, where $Q=0,\pm 1,\pm 2, ...$ 
is the topological charge and
 $N_{\pi^+}$ is the positive pion multiplicity in 
one unit of rapidity.
More accurate estimates~\cite{Kharzeev:2007jp}
were found to be close to
 the same number, of the order of $10^{-2}$ for mid-central collisions,
which corresponds to $10^{-4}$ for the correlator $\mean{a_\alpha a_\beta}$
(third term in Eq.~\ref{eq:v-a}).
Without in medium effects, one expects the symmetry between same- and opposite-sign correlations:
$\mean{a_+a_+}=\mean{a_-a_-}=-\mean{a_+a_-} >0$,
but in case of heavy ion collisions
one needs to account for correlation modification due to
particle interaction with the medium.
This is similar to the
back-to-back suppression of jet-like correlations,
and is predicted to result in in suppression of opposite sign correlations~\cite{Kharzeev:2007jp}:
$\mean{a_+a_+}=\mean{a_-a_-} \gg -\mean{a_+a_-}$. 
The correlator $\mean{a_\alpha a_\beta}$
should follow (a typical for any
kind of correlations due to clusters)
$1/N_{ch}$ dependence ($N_{ch}$ is the charge multiplicity).
Local parity violation 
is a non-perturbative effect and
the main contribution should come
from particles which have transverse momentum smaller than
1~GeV/$c$~\cite{Kharzeev:2007jp},
with correlation width of the order of one unit
in rapidity (typical size of the \P-odd domain is about 1~fm).

\section{Measurement technique}

The reaction plane orientation is not known a priori in the experiment,
and it needs to be reconstructed from particle azimuthal 
distributions~\cite{Poskanzer:1998yz}.
For that reason instead of measuring the two particle
correlator wrt. the reaction plane (\ref{eq:obs1}),
we evaluate the three particle
correlator, where the third particle, $c$,
gives an estimate of the reaction plane.
Such a three particle correlator can be related to the
correlator (\ref{eq:obs1}) by the following equation:
\be
\la \cos(\phi_a +\phi_\beta -2\psirp) \ra 
=
\la \cos(\phi_a +\phi_\beta -2\phi_c) \ra / v_{2,c}.
\label{e3p}
\ee
Here $v_{2,c}$ is an elliptic flow value of the third particle $c$,
and it accounts for the finite resolution in our estimate of the reaction plane angle
\cite{Poskanzer:1998yz,Borghini:2001vi,Adams:2003zg}.
Equation \ref{e3p} assumes that particle $c$ is correlated with particles
$\alpha$ and $\beta$ only via common correlation to the reaction plane.
This third particle further complicates experimental data analysis,
since it potentially adds 
to the measured values of the correlator~(\ref{e3p})
some background contributions from genuine 3-particle correlations.
These contributions are discussed in Sec.~\ref{PhysicsBackgrounds}.

\section{Experimental setup}
\label{sec:setup}

The results presented here are based on 14.7M Au+Au
and 13.9M Cu+Cu collisions at the incident energies
\snn=200~GeV, and 2.4M Au+Au and 6.3M Cu+Cu events at \snn=62~GeV. 
Collisions were recorded with the STAR detector
during the 2004 and 2005 runs. 
Charged particle tracks were reconstructed in a cylindrical
and azimuthally symmetric
Time Projection Chamber (TPC)~\cite{Ackermann:2002ad,Anderson:2003ur}. 
A minimum bias trigger was used during data-taking
with offline collision vertex cut of 30~cm along the beam line
from  the center of the main TPC.

The correlations are reported in the pseudorapidity region $|\eta| < 1.0$
with particle momentum $0.15 < p_t < 2.0$~GeV/c
(unless stated otherwise).
Standard STAR track quality cuts are applied \cite{Abelev:2009tx}.
The Large elliptic flow measured at
RHIC~\cite{Ackermann:2000tr} is used to estimate the reaction plane 
from particle distributions in the main TPC
and two Forward Time Projection Chambers (FTPC)~\cite{Ackermann:2002yx}.
The latter cover pseudorapidity intervals $2.7 < | \eta | < 3.9$.
In the most forward direction, STAR has two Zero Degree 
Calorimeter - Shower Maximum Detectors 
(ZDC-SMD)~\cite{Adams:2005ca,Adler:2001fq}.
ZDC-SMDs are sensitive 
to the directed flow of neutrons in the beam rapidity regions,
which is used for the reaction plane reconstruction.

\section{Understanding detector effects}
\label{sec:detector_effects}

Detector effects have been corrected with recentering 
procedure~\cite{Poskanzer:1998yz}.
Corrections are applied run-by-run, separately
for positive and negative particles, 
for each centrality bin, and as a function
of particle pseudorapidity and transverse momentum.
The validity of the recentering method was verified
by calculating three-particle cumulants according 
to~\cite{Borghini:2002vp,Selyuzhenkov:2007zi}.

\begin{figure}[h]
\includegraphics[width=.49\textwidth]{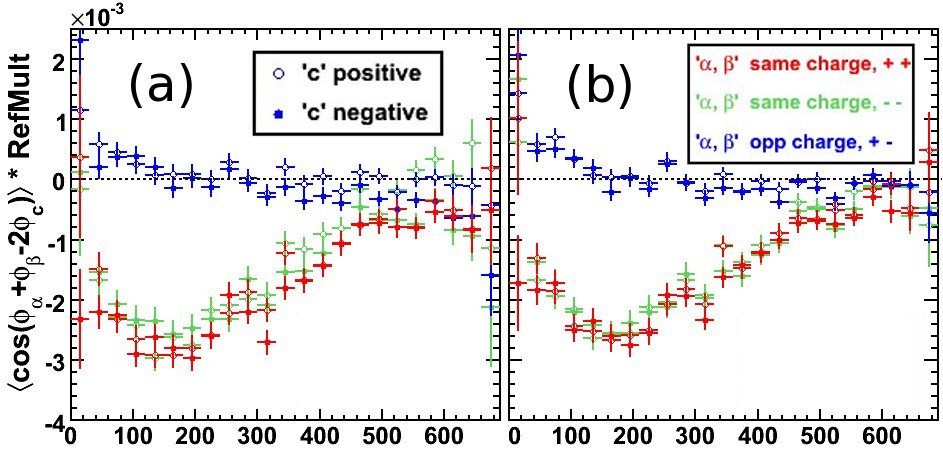}
\centerline{~~~~~~~~RefMult~~~~~~~~~~~~~~~~~~~~~~~~~~~RefMult}
  \caption{ $\mean{\cos(\phia +\phib -2\phi_c)}$ as a function of reference
  multiplicity for different charge combinations~\cite{Abelev:2009tx}: (a) before
corrections for acceptance effects, and (b) after corrections.
In the legend the signs indicate the charge of particles
$\alpha$,~$\beta$,~and~$c$. 
}
  \label{fig:no_recenter}
  \label{fig:recenter}
\end{figure}

Figure~\ref{fig:no_recenter} shows three-particle correlator 
$\la \cos(\phi_a +\phi_\beta -2\phi_c) \ra$ as a function
of reference multiplicity (measured charge multiplicity in $|\eta|<1$)
in Au+Au collisions at \snn=200~GeV
(a) before and (b) after correcting for detector effects.
The signal is scaled by the reference multiplicity to better
indicate the acceptance effects for central collisions.
All the differences which do not depend on the relative charge of $\alpha$ and $\beta$
disappear after the acceptance correction.
We have performed several additional checks to ensure that the signal is
not due to detector effects:
 \begin{itemize}
 \item
Distortions in the track momenta due to the charge buildup
in the STAR Time Projection Chamber (TPC) at high accelerator luminosity.
Results obtained from
RHIC Run II (a low luminosity run), and those from
Run IV divided into high and low luminosity events,
yield the same signal within statistical uncertainties. 
 \item
Signal dependence on reconstructed position of the collision vertex along the beam line
has been checked, and no dependence has been found.
 \item
Displacement of track hits when it passes the TPC central membrane.
The correlator (\ref{e3p}) has been calculated using only particles with the entire track
in one of the different half-barrels of the TPC.
Corrected for the  signal  dependence  on  the  track  
separation in pseudorapidity, results are found to be 
consistent with those obtained before introducing the rapidity separation.
 \item  
Feed-down effects from non-primary tracks (i.e. resonance decay daughters)
have been studied via cuts on track distance of closet approach ($dca$).
Results for $dca<1$~cm and $dca<3$~cm
are found to be consistent within statistical errors.
 \item
Electron contribution to the measured signal has been
checked via specific energy loss ($dE/dx$) in the
volume of the TPC and found to be negligible.
This check was done to test whether the signal was due to hadron production, or
lepton production.
 \item
Studied a correlator similar to (\ref{e3p})
but with the reaction plane angle rotated by $\pi/4$.
This new correlator should only deviate from zero due to detector effects.  It was
found to be consistent with zero within statistical errors.
 \item
Variation depending on the charge of the third particle used to reconstruct the reaction plane
and changes of the STAR magnetic field polarity.
The variations does not change the observed signal.
\end{itemize}

Our conclusion from a number of tests performed
is that detector effects are not responsible
for the observed correlations.
\section{Testing sensitivity to 2-particle correlations wrt. the reaction plane}
\label{test2partCorr}
 
\begin{figure}[ht]
  \includegraphics[width=0.49\textwidth]{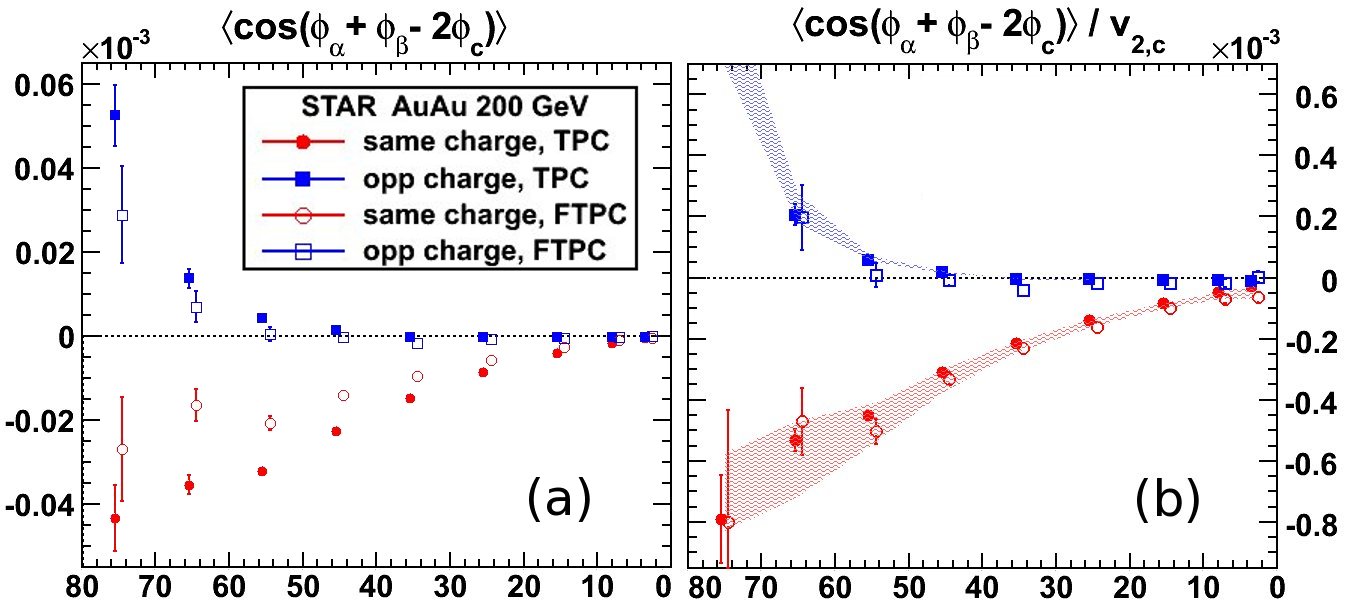}
\centerline{~~~~~~~~~~~~~~~~\% Most central~~~~~~~~~~~~~~~~~\% Most central}
  \caption{
    (a) $\mean{\cos(\phia +\phib -2\phi_c)}$ obtained by
    selecting the  third particle  from the main TPC
    and  the Forward  TPCs~\cite{Abelev:2009tx}.
    (b) $\mean{\cos(\phia +\phib -2\phi_c)}$ after scaling by the flow of the  third  particle.
The shaded areas represent the systematic uncertainty from $v_{2,c}$ scaling.
}
  \label{fig:au200F}
\end{figure}

Figure~\ref{fig:au200F}(a)    compares  the    three-particle
correlations  $\mean{\cos(\phia +\phib -2\phi_c)}$ obtained for different charge combinations, as a function
of centrality, when the  third particle is
selected  from the  main  TPC (solid symbols) with when it is selected
from  the Forward TPCs (open symbols).
Elliptic  flow  data have  been  taken  from Ref.~\cite{Adams:2004bi,Voloshin:2007af}.             

Assuming  that the second  harmonic of  the  third particle  is
correlated with  the first harmonic  of the first  two particles
via  a  common  correlation  to the  reaction  plane,  the
correlator should  then be proportional to the  elliptic flow of
the third particle.   
Figure~\ref{fig:au200F}(b) shows very good agreement between the same
charge correlations obtained with the third particle in the TPC and
FTPC regions, which supports for such correlations the assumption of Eq.~\ref{e3p}.
The opposite charge correlations are small in magnitude and 
with current statistics it is difficult to conclude
on validity of the same assumption. 
Results obtained with the  event plane reconstructed with
ZDC-SMD are consistent with those shown in Fig.~\ref{fig:au200F}(b),
though the statistical errors  on ZDC-SMD results are about 5 times larger. 

Agreement between TPC, FTPC, and ZDC SMD results
give us a confidence that 3-particle correlation
that we measure are indeed sensitive
to the 2-particle correlations wrt. the reaction plane.
\section{Modeling physics backgrounds}
\label{PhysicsBackgrounds}

\begin{figure}[ht]
 \includegraphics[width=.42\textwidth]{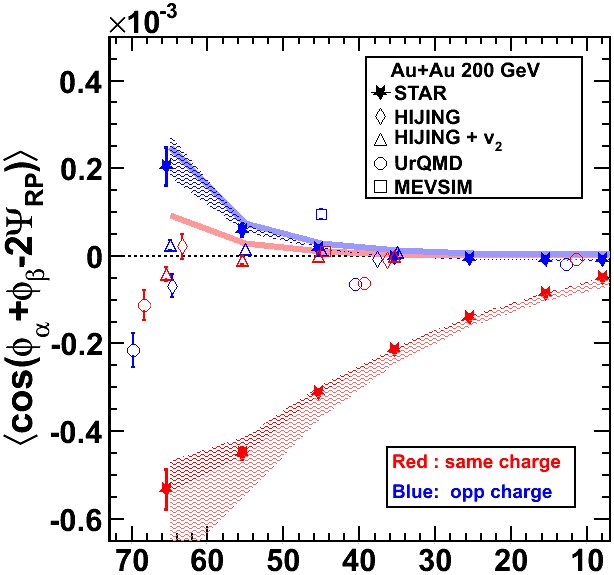}

\vspace{0.1cm}
\centerline{\Large ~~~~~~~~~~~~~~~~~~~~~~~~\% Most central}
 \caption{
Comparing correlator (\ref{e3p})
from STAR data~\cite{Abelev:2009tx} vs. various physics background event generators
in Au+Au collisions at $\sqrt{s_{NN}}=200$~GeV.
None of the generators can reproduce same-charge correlations.
}
 \label{fig:AuAusimulations}
\end{figure}

Figure~\ref{fig:AuAusimulations}
shows the correlator (\ref{e3p}) for Au+Au
collisions at \snn=200~GeV.
Blue symbols and lines mark opposite-charge
correlations, and red are same-charge.  
Star symbols represent the STAR data,
with shaded area reflecting the uncertainty due to $v_2$ scaling.
Other symbols in Fig.~\ref{fig:AuAusimulations} show
possible background two particle correlations wrt. the reaction plane
estimated from various Monte-Carlo event generator
with true reaction plane:
\begin{itemize}
\item
HIJING~\cite{refHIJING} with default, quenching-off setting:
with (triangles), and without (diamonds) $v_2$-afterburner.
Elliptic flow is added by "shifting" method~\cite{Poskanzer:1998yz},
with $v_2$ values consistent with STAR measurements at the given centrality.
\item
UrQMD~\cite{refRQMD} with default setting (circles).
\item
MEVSIM~\cite{refMEVSIM} Monte-Carlo simulations (squares).
MEVSIM tests resonance ($\phi$, $\Delta$, $\rho$, $\omega$, and $K^{*}$) contribution with realistic elliptic flow pattern.
\end{itemize}
Thick solid lines in Fig.~\ref{fig:AuAusimulations} indicate genuine three particle correlations from HIJING
(corresponding estimates from UrQMD are about factor of two smaller).
HIJING three particle correlations produce data-like opposite-sign signal
(compare blue line with blue stars in Fig.~\ref{fig:AuAusimulations}),
which may indicate dilution of the  opposite-sign signal
with effects not related to the reaction plane orientation.

Figure~\ref{fig:AuAusimulations} shows
that no generator gives qualitative agreement with the data,
thought they produce non-zero correlations.
These models do not match the correlations for
${\mean{\cos(\phia-\phib)}}$ that are seen in the data either,
which points to the
need for better modeling of two-particle correlations to give
quantitatively meaningful comparisons for correlator (\ref{e3p}).
Other background effects studied
are (see for details~\cite{Abelev:2009tx}):
\begin{itemize}
\item
Correlations from processes in which particles $\alpha$ and
$\beta$ are products of a cluster (e.g. resonance, jet,
 di-jets) decay, and the cluster itself exhibits elliptic
flow or decays (fragments)
differently when emitted in-plane compared to out-of-plane
\item
Jets  as potential source of reaction-plane dependent
background since their properties
may vary with respect to the reaction plane.
\item
Cluster formation, which plays an important role in multiparticle
production at high energies,
and may account for production of a significant fraction of all particles.
\item
Directed flow, which on average is zero in a symmetric pseudorapidity
interval, but can contribute to the correlator (\ref{e3p}) via 
flow fluctuations (see the first term in Eq.~\ref{eq:v-a}). 
\item
Effect of global polarization (discussed in the introduction),
which produce polarized secondary particles along the direction of
the system's angular momentum.
\end{itemize}
From our studies we conclude that
none of the background effects listed above
can be responsible for the same-sign correlations shown in Fig. \ref{fig:AuAusimulations}.

\section{Results}
\label{sec:results}
\begin{figure}[ht]
 \includegraphics[width=.44\textwidth]{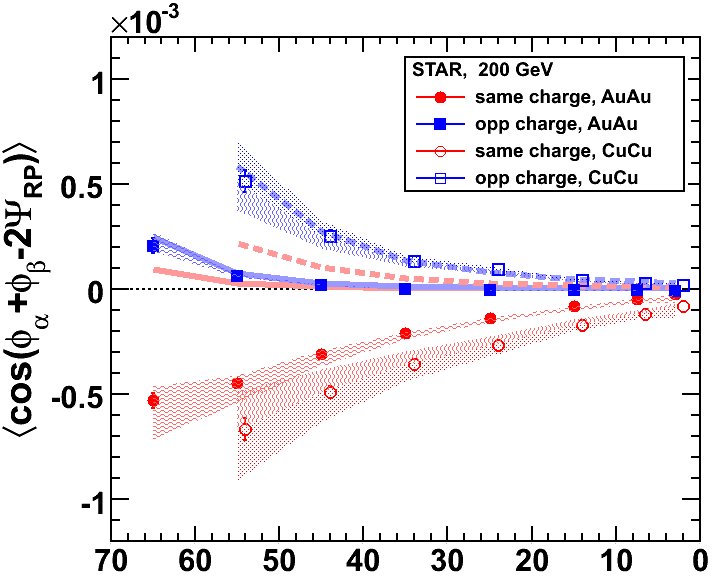}
\vspace{0.1cm}
\centerline{\Large ~~~~~~~~~~~~~~~~~~~~~~~~~~\% Most central}
 \caption{ Correlator (\ref{e3p}) in Au+Au and Cu+Cu
collisions at $\sqrt{s_{NN}}=200$~GeV ~\cite{Abelev:2009tx}.
The shaded band indicates uncertainty associated with $v_2$ scaling.
Thick solid (Au+Au) and dashed (Cu+Cu) 
lines represent HIJING estimates
for possible non-reaction-plane dependent contribution from
many-particle correlations.
}
 \label{fig:uuv2_200}
\end{figure}

Figure~\ref{fig:uuv2_200} shows correlator (\ref{e3p}) in Au+Au and Cu+Cu
collisions at $\sqrt{s_{NN}}=200$~GeV.
The correlations are weaker in more central collisions compared to more
peripheral collisions, which partially can 
be attributed to dilution of correlations
which occurs in the case of particle production from multiple sources. 
The correlations in Cu+Cu collisions (open symbols),
appear to be larger than the correlations in Au+Au (solid symbols)
for the same centrality of the collision.
One reason for this difference may be the difference in number of
participants (or charge multiplicity) in Au+Au and Cu+Cu collisions 
at the same centrality. 
The local \P-violation signal is expected to have approximately $1/N_{ch}$ dependence, and at 
the same centrality of the collision the multiplicity is smaller 
in Cu+Cu collisions compared to Au+Au.
The difference in magnitude between same and opposite charge correlations
is considerably smaller in Cu+Cu than in Au+Au,
qualitatively in agreement with the scenario of stronger suppression 
of the back-to-back correlations in Au+Au collisions.

\begin{figure}[ht]

 \includegraphics[width=.49\textwidth]{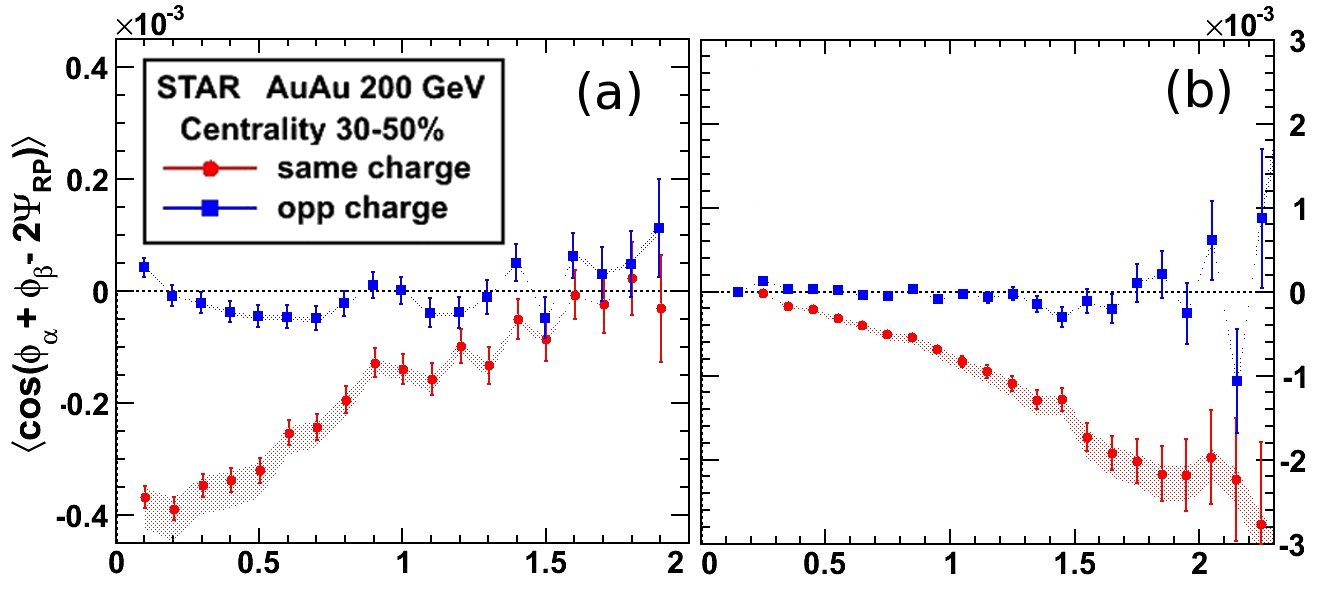}
\centerline{~~~~~~~~~~~~~~~~$|\eta_\alpha-\eta_\beta|$ ~~~~~~~~~~~~~~~~($p_{t,\alpha}+p_{t,\beta})/2$, GeV/$c$ }
 \caption{Au+Au at 200 GeV~\cite{Abelev:2009tx}.
The correlations dependence on
(a) pseudorapidity separation,
$|\eta_\alpha -\eta_\beta|$,
and (b) on $(p_{t,\alpha}+p_{t,\beta})/2$
for centrality 30-50\%.
The shaded band indicates uncertainty associated with $v_2$ scaling.
}
 \label{fig:uuv2pt}
\end{figure}

Figure~\ref{fig:uuv2pt}(a) shows the dependence of the correlator~(\ref{e3p})
on the difference in pseudorapidities of two particles,
 $|\eta_\alpha -\eta_\beta|$, for 30-50\% centralities,
while Fig.~\ref{fig:uuv2pt}(b)
shows its dependence on the sum of
the transverse momentum magnitudes (without upper $p_t$ cut)
for the same centrality region.
The correlations have a typical hadronic width of about one unit of
pseudorapidity, but we do not observe them
to be concentrated in the low $p_t$
region as naively might be expected for \P-violation effects. 

Other results, in particular for collisions at \snn=62.4~GeV,
correlation scaling with number of participants,
and their dependence on transverse momentum difference,
can be found in \cite{Abelev:2009tx}.

\section{Summary}

Formation of local \P-odd domains has been predicted in
nuclear collisions \cite{Kharzeev:2004ey}.
This effect should result in charge separation along the orbital momentum
of the system created in non-central heavy-ion collisions.
The STAR measurements of 3-particle azimuthal correlations
in Au+Au and Cu+Cu collisions at $\sqrt{s_{NN}}$=200 and 62~GeV
reveal non-zero signal.
Measured 3-particle azimuthal correlations
are directly sensitive to local strong parity violation,
but susceptible to contributions from \P-conserving backgrounds.
So far we could not explain the observed same sign correlations,
and data can not be describe with any of the existing Monte-Carlo event generators
of the heavy-ion collisions.
At the same time, qualitatively the data agrees with predictions for local \P-violation
(though the signal persists to higher transverse momentum than expected).
The presented measurements demand detailed theoretical calculations
of the \P-violating signal and backgrounds
to make a definitive statement about possible local strong \P-violation in 
heavy ion collisions.



\medskip
\noindent
\begin{thebibliography}{99}

\bibitem{Kharzeev:2004ey}
  D.~Kharzeev,
  Phys.\ Lett.\  B {\bf 633}, 260 (2006).

\bibitem{Kharzeev:2007jp}
  D.~E.~Kharzeev, L.~D.~McLerran and H.~J.~Warringa,
  Nucl.\ Phys.\  A {\bf 803}, 227 (2008).

\bibitem{Fukushima:2008xe}
  K.~Fukushima, D.~E.~Kharzeev and H.~J.~Warringa,
  Phys.\ Rev.\  D {\bf 78}, 074033 (2008).

\bibitem{Liang:2004ph}
  Z.~T.~Liang and X.~N.~Wang,
  Phys.\ Rev.\ Lett.\  {\bf 94}, 102301 (2005)
  [Erratum-ibid.\  {\bf 96}, 039901 (2006)].

\bibitem{Gao:2007bc}
  J.~H.~Gao, S.~W.~Chen, W.~t.~Deng, Z.~T.~Liang, Q.~Wang and X.~N.~Wang,
  Phys.\ Rev.\  C {\bf 77}, 044902 (2008).

\bibitem{Voloshin:2008dg}
  S.~A.~Voloshin, A.~M.~Poskanzer and R.~Snellings,
  arXiv:0809.2949 [nucl-ex].


\bibitem{Voloshin:2004ha}
  S.~A.~Voloshin,
  arXiv:nucl-th/0410089.


\bibitem{Abelev:2007zk}
  B.~I.~Abelev {\it et al.}  [STAR Collaboration],
  Phys.\ Rev.\  C {\bf 76}, 024915 (2007).

\bibitem{Abelev:2008ica}
  B.~I.~Abelev {\it et al.}  [STAR Collaboration],
  Phys.\ Rev.\  C {\bf 77}, 061902 (2008).

\bibitem{Diakonov:2009jq}
  D.~Diakonov,
  arXiv:0906.2456 [hep-ph].



\bibitem{Pospelov:1999ha}
  M.~Pospelov and A.~Ritz,
  Phys.\ Rev.\ Lett.\  {\bf 83}, 2526 (1999).

\bibitem{Baker:2006ts}
  C.~A.~Baker {\it et al.},
  Phys.\ Rev.\ Lett.\  {\bf 97}, 131801 (2006).

\bibitem{Voloshin:2004vk}
  S.~A.~Voloshin,
  Phys.\ Rev.\  C {\bf 70}, 057901 (2004).

\bibitem{Borghini:2002vp}
 N.~Borghini, P.~M.~Dinh and J.~Y.~Ollitrault,
 Phys.\ Rev.\  C {\bf 66}, 014905 (2002).

\bibitem{Adams:2003zg}
  J.~Adams {\it et al.}  [STAR Collaboration],
  Phys.\ Rev.\ Lett.\  {\bf 92}, 062301 (2004).

\bibitem{Poskanzer:1998yz}
  A.~M.~Poskanzer and S.~A.~Voloshin,
  Phys.\ Rev.\  C {\bf 58}, 1671 (1998).

\bibitem{Borghini:2001vi}
  N.~Borghini, P.~M.~Dinh and J.~Y.~Ollitrault,
  Phys.\ Rev.\  C {\bf 64}, 054901 (2001).

\bibitem{Ackermann:2002ad}
  K.~H.~Ackermann {\it et al.}  [STAR Collaboration],
  Nucl.\ Instrum.\ Meth.\  A {\bf 499}, 624 (2003).

\bibitem{Anderson:2003ur}
  M.~Anderson {\it et al.},
  Nucl.\ Instrum.\ Meth.\  A {\bf 499}, 659 (2003).

\bibitem{Abelev:2009tx}
\mbox{B.~I.~Abelev {\it et al.} [STAR Collaboration]},
  arXiv:0909.1717 [nucl-ex].

 \bibitem{Ackermann:2000tr}
    K.~H.~Ackermann {\it et al.}  [STAR Collaboration],
    Phys.\ Rev.\ Lett.\  {\bf 86}, 402 (2001).


\bibitem{Ackermann:2002yx}
  K.~H.~Ackermann {\it et al.},
  Nucl.\ Instrum.\ Meth.\  A {\bf 499}, 713 (2003).

\bibitem{Adams:2005ca}
  J.~Adams {\it et al.}  [STAR Collaboration],
  Phys.\ Rev.\  C {\bf 73}, 034903 (2006).

\bibitem{Adler:2001fq}
  C.~Adler, {\it et al.},
  Nucl.\ Instrum.\ Meth.\  A {\bf 461}, 337 (2001);
STAR ZDC-SMD proposal, STAR Note SN-0448
(2003).

\bibitem{Selyuzhenkov:2007zi}
 I.~Selyuzhenkov and S.~Voloshin,
 Phys.\ Rev.\  C {\bf 77}, 034904 (2008).

\bibitem{Adams:2004bi}
  J.~Adams {\it et al.}  [STAR Collaboration],
  Phys.\ Rev.\  C {\bf 72}, 014904 (2005).

\bibitem{Voloshin:2007af}
 S.~A.~Voloshin  [STAR Collaboration],
  J.\ Phys.\ G {\bf 34}, S883 (2007).

\bibitem{refHIJING} 
 M. Gyulassy and X.-N. Wang, Comput. Phys. Commun. {\bf 83},
  307 (1994);
  X.N. Wang and M. Gyulassy, Phys. Rev. D {\bf 44}, 3501 (1991).

\bibitem{refRQMD} 
  S.~A.~Bass {\it et al.},
  Prog.\ Part.\ Nucl.\ Phys.\  {\bf 41}, 255 (1998)
  [Prog.\ Part.\ Nucl.\ Phys.\  {\bf 41}, 225 (1998)].

\bibitem{refMEVSIM}
  R.~L.~Ray and R.~S.~Longacre,
  arXiv:nucl-ex/0008009.


\end{thebibliography}
\end{document}